\documentclass[]{interact}

\usepackage{epstopdf}% To incorporate .eps illustrations using PDFLaTeX, etc.
\usepackage[caption=false]{subfig}% Support for small, `sub' figures and tables
\usepackage{braket}
\usepackage{amsmath, amssymb, amsfonts}

\usepackage[numbers,sort&compress]{natbib}% Citation support using natbib.sty
\bibpunct[, ]{[}{]}{,}{n}{,}{,}% Citation support using natbib.sty
\renewcommand\bibfont{\fontsize{10}{12}\selectfont}% Bibliography support using natbib.sty
\makeatletter% @ becomes a letter
\def\NAT@def@citea{\def\@citea{\NAT@separator}}% Suppress spaces between citations using natbib.sty
\makeatother% @ becomes a symbol again

\theoremstyle{plain}% Theorem-like structures provided by amsthm.sty

\theoremstyle{definition}

\theoremstyle{remark}

\usepackage{listings,jvlisting}
\usepackage{url}

\usepackage{xspace}
\newcommand{\aenet}{\textrm{\ae{}net}\xspace}
\newcommand{\conf}{\vec{\sigma}}

\usepackage[version=4]{mhchem}

\begin{document}

%article types: Review, Research Article, Rapid Communication, Case Report, Tutorial,  Introductory Review, and Data Note
\articletype{}% Specify the article type or omit as appropriate

\title{Configuration sampling in multi-component multi-sublattice systems enabled by ab Initio Configuration Sampling Toolkit (abICS)}

\author{
\name{Shusuke Kasamatsu\textsuperscript{a}\thanks{CONTACT Shusuke Kasamatsu Email: kasamatsu@sci.kj.yamagata-u.ac.jp}, Yuichi Motoyama\textsuperscript{b}, Kazuyoshi Yoshimi\textsuperscript{b}, and Tatsumi Aoyama\textsuperscript{b}}
\affil{\textsuperscript{a}Academic Assembly (Faculty of Science), Yamagata University, 1-4-12 Kojirakawa, Yamagata-shi, Yamagata 990-8560 Japan; \textsuperscript{b}Institute for Solid State Physics, University of Tokyo, Chiba 277-8581, Japan}
}

\maketitle

\begin{abstract}
Simulation of the intermediate levels of disorder found in multi-component multi-sublattice systems in various functional materials is a challenging issue, even for state-of-the-art methodologies based on first-principles calculation. Here, we introduce our open-source package ab Initio Configuration Sampling Toolkit (abICS), which combines high-throughput first-principles calculations, machine learning, and parallel extended ensemble sampling in an active learning setting to enable such simulations.
The theoretical background is reviewed in some detail followed by brief notes on usage of the software. In addition, our recent applications of abICS to multi-component ionic systems and their interfaces for energy applications are reviewed as demonstration of the power of this approach.
\end{abstract}


\section{Introduction}
Inorganic crystalline materials for applications including energy conversion/storage, catalysis, nanoelectronics, and structural materials 
always contain some amount of impurities, lattice defects, and configurational disorder that depend on the processing conditions. Very few `pure' materials are used in actual applications; rather, introduction of impurities and/or disorder is employed almost ubiquitously for improving the properties and functionalities of various materials. For example, insulating materials are turned into ion conductors for solid state batteries and fuel cells through impurity doping and defect engineering \cite{takeiri2022,jclement2020,hyodo_fast_nodate,hyodo_accelerated_2021}, and high-entropy (i.e., disordered) alloys are under intense research for magnet and structural applications \cite{fukushima2022}. Control of
atomistic configurations is of interest also for phonon engineering \cite{hu2020}.

When the material can be considered a random alloy, the coherent potential approximation (CPA) \cite{shiba_reformulation_1971,soven_application_1970,akai_fast_1989} combined with the Korringa-Kohn-Rostoker method \cite{KORRINGA1947392,kohn_solution_1954} is a very fast and powerful electronic structure method for properties prediction from first principles. Another approach that can be used in standard density functional theory codes is the supercell approach based on special quasi-random structures \cite{zunger_special_1990,van_de_walle_efficient_2013}. These methods enable single-shot calculations of a completely random alloy.

On the other hand, complex ionic solids often show strong defect/impurity association due to Coulomb interactions between ions with varying charges. This means that there is some amount of `order within disorder', and the assumption of a completely random alloy breaks down even at sintering temperatures nearing 2000 K \cite{Kasamatsu2020}. Moreover, it has been pointed out recently that significant levels of short-range order exists even in high-entropy alloys with a profound effect on the thermodynamics and the resulting magnetic and mechanical properties \cite{ferrari_frontiers_2020}. 
In these cases, thermodynamically relevant lattice configurations need to be elucidated before properties prediction. 
An exhaustive calculation over all possible configurations is impossible except when impurities and defects are very dilute, and in such cases, various importance sampling methods are employed. 
First-principles molecular dynamics can be used for sampling structures of quickly-relaxing (i.e., liquid) systems, but it is nearly useless for most solid-state systems with high energy barriers between different configurations. On the other hand, Monte Carlo (MC) algorithms are not bound by the physical relaxation times because they can employ unphysical trial steps such as the swapping of atom positions. Moreover, in systems that can be mapped onto a lattice, it is often sufficient to sample configurations that are relaxed starting from an ideal lattice with varying occupations of each site by different atomic species or vacancies. Thus, the natural way to go would be to combine MC simulations with first-principles relaxation and energy calculations. However, because of the high computational cost of first-principles calculations, such reports are still very few \cite{Khan2016, Kasamatsu_2019, Wexler2019, Kasamatsu2020, Fujii2021}; instead, the standard way is to fit a cluster expansion Hamiltonian to first-principles results on a limited number of configurations and then use that effective model for MC sampling \cite{Sanchez1984,Ceder1993,Walle2002,Seko2009,Sanchez2010,Sanchez2017,Wu2016,chang2019}. 

Although cluster expansion has been very successful, its application to many-component multi-sublattice ionic systems is often impractical due to combinatorial explosion in the necessary number of terms (i.e., clusters) in the expansion (see Ref.~\cite{barroso-luque_cluster_2022} for a recent reformulation and review of the cluster expansion method with an emphasis on application to multicomponent ionic materials). To deal with this, we proposed to use more flexible non-linear models, i.e., machine-learning potentials that have seen explosive development in recent years for enabling molecular dynamics simulations with near-first principles accuracy at a small fraction of the cost. The nonlinearity of the model should, in principle, enable higher accuracy with less complex descriptors than linear models such as cluster expansion.
Moreover, machine learning potentials usually map continuous coordinates to continuous energies, but we showed that it can map coordinates on a lattice \emph{before} relaxation to the total energy \emph{after} relaxation \cite{Kasamatsu2022}, which is the only information needed for thermodynamic sampling on a lattice. This affords a speedup by a factor of 10 to a few hundred compared to performing relaxations on each configuration encountered during MC sampling. 

Based on the above ideas, we have been developing an open source framework abICS ({\it ab Initio} Configuration Sampling), which facilitates the combination of high-throughput first-principles calculation, training of the machine learning configuration energy model, parallel extended-ensemble MC sampling using the trained model, and iterative improvement of the model through active learning cycles. In this Review, we report the basic algorithms and features implemented in abICS, some details of the installation and usage of the code, and a few example applications. Finally, we will conclude with some comments on future development of abICS.

\section{Active learning workflow using abICS}
\begin{figure}[t]
    \centering
    \includegraphics[width=14cm]{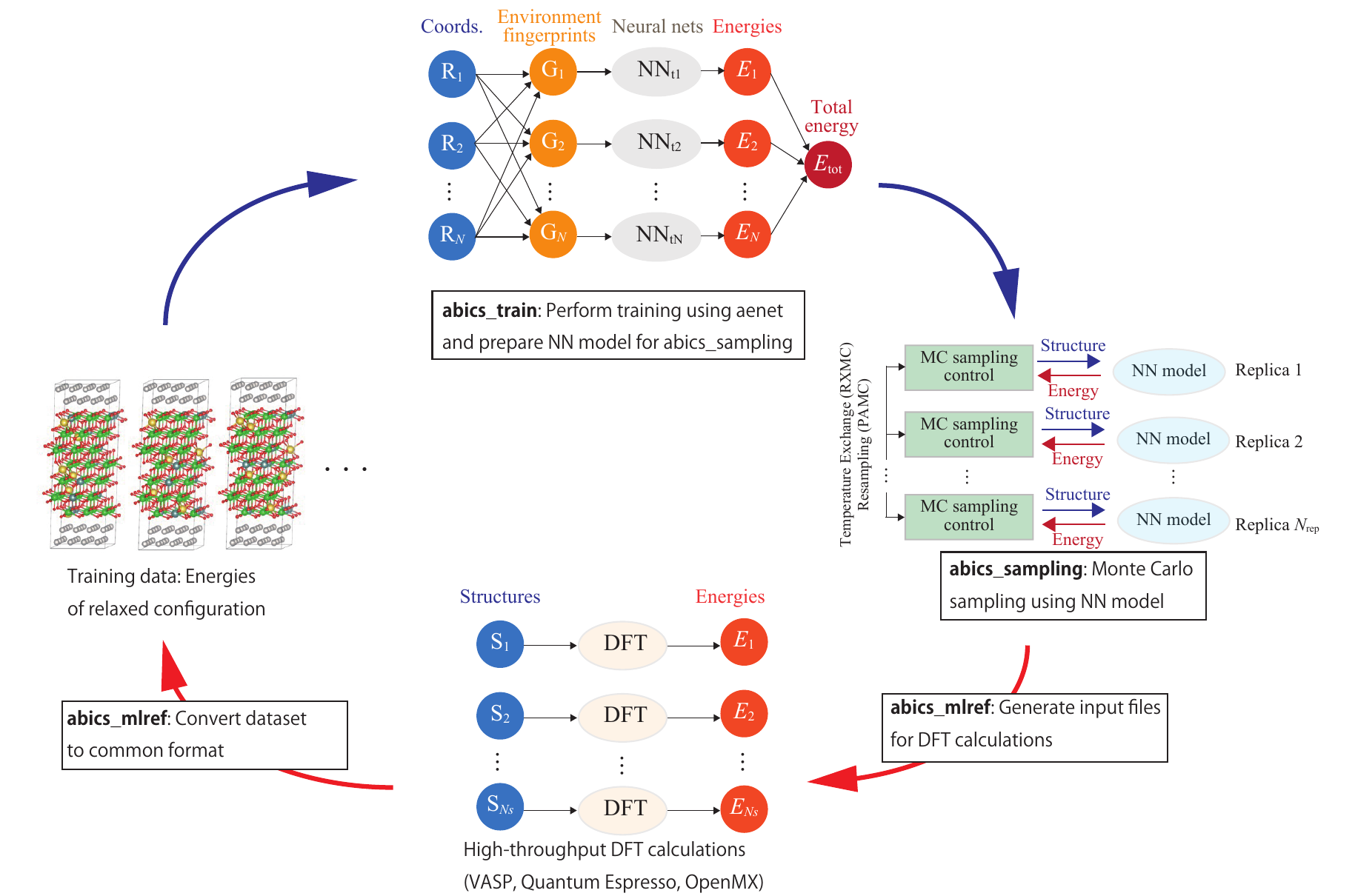}
    \caption{A schematic of the framework implemented in abICS.}
    \label{fig:schematic}
\end{figure}

The workflow of the active learning method implemented in abICS is shown in Fig.~\ref{fig:schematic}. A typical procedure is outlined as follows, which involves Python programs \texttt{abics\_mlref}, \texttt{abics\_train}, and \texttt{abics\_sampling} which are part of the abICS package.

\subsection{Generation of initial training dataset of random configurations (\texttt{\upshape{abics\_mlref}})}
\texttt{abics\_mlref} is used to generate random configurations on an ideal lattice and prepare input files for first-principles relaxation and energy calculations using VASP \cite{Kresse1996}, Quantum ESPRESSO \cite{giannozzi_advanced_2017}, or OpenMX \cite{ozaki_variationally_2003}. The user is tasked with setting up how to run the first-principles calculations: how many CPU cores (or even GPUs) to be used for each configuration, how many configurations are calculated in parallel, adapting job submission scripts for their supercomputer center to enable high-throughput calculations, etc. After finishing the calculations, \texttt{abics\_mlref} is rerun to convert the results into a common format independent of the first-principles calculation code.

\subsection{Neural network Training (\texttt{\upshape{abics\_train}})} \label{sec:nntrain}
   After conversion of the first-principles calculation results using \texttt{abics\_mlref}, \texttt{abics\_train} is used to train the configuration energy model on those results.
    Currently, \texttt{abics\_train} interfaces with \aenet\cite{artrith2016,artrith2017} for training a neural network model.

\subsection{Monte Carlo sampling (\texttt{\upshape{abics\_sampling}})}
    Next, \texttt{abics\_sampling} is used to perform replica exchange Monte Carlo (RXMC) sampling \cite{Hukushima1996} or population annealing Monte Carlo (PAMC) sampling \cite{Hukushima2003} using the trained neural network configuration energy model (the algorithms are explained in Sec.~\ref{sec:MC}). \texttt{abics\_sampling} can perform canonical sampling, which maintains a fixed number of atoms, as well as grand canonical sampling, which allows for changes in the composition. The latter is still in the testing phase.
    %Running the program creates directories named by the replica numbers under the current directory, and each replica runs the solver in it.
    At each MC sampling step, \texttt{abics\_sampling} invokes a model calculator (only \aenet is supported as of present)  to evaluate the energy of sampled configurations.

\subsection{Validation and iterative refinement (\texttt{\upshape{abics\_mlref}})}
    After MC sampling, \texttt{abics\_mlref} is used to take
    a subset of configuration samples from the MC results and produce corresponding input files for first-principles calculations. The user runs these calculations, then \texttt{abics\_mlref} is run again to convert the results to a common format to be read by \texttt{abics\_train}. At the same time, a plot that compares
    the configuration energy model to the first-principles results are produced, which the user can use for validation of the model accuracy. If the accuracy is insufficient, the data is added to the training data set, and the procedure is repeated from the training step (Sec.~\ref{sec:nntrain}). This iterative active learning process refines the neural network model, improving its accuracy with the number of iterations.

\subsection{Production Monte Carlo run (\texttt{\upshape{abics\_sampling}})}
     Once a neural network configuration energy model with sufficient accuracy is obtained, it is recommended to perform a production MC run using an expanded supercell and/or a larger number of sampling steps. \texttt{abics\_sampling} can be used for this task. Generally, we have found that the number of MC steps needed for obtaining converged statistics (e.g., short-range or intermediate-range order parameters, similarity measures, ion concentration profiles, etc.) is much larger than that necessary for generating sufficient training data during the active learning iterations.

\section{Algorithms}
In this section, we describe the main algorithms implemented in abICS.

\subsection{On-lattice configuration energy model} \label{sec:olmodel}
As mentioned above, abICS relies, at present, on \aenet code\cite{artrith2016,artrith2017} to train the energy model. \aenet implements the high-dimensional neural network potential (NNP) originally proposed by Behler and Parinello \cite{behler2007,behler2011,behler2015}, which assumes that the total energy can be expressed as a sum of atomic energies that are determined by the environment around each atom and uses neural networks to fit these atomic energies. The same network is used for the same atomic species, so the total energy is  expressed as
\begin{equation}\label{eq:NNP}
    E(\conf) = \sum_i^\text{atoms} \text{NN}_{t_i} (f[\conf_i^{R_\text{c}}]),
\end{equation}
where $\conf_i^{R_\mathrm{c}}$ represents the environment around atom $i$ within a finite cutoff distance $R_\mathrm{c}$, $f$ is a fingerprint function that maps the atomic environment to a multidimensional descriptor vector, $t_i$ is the species type of atom $i$, and 
$\mathrm{NN}_{t_i}$ denotes the neural network for 
species $t_i$ that maps the environment descriptor vector to an atomic energy. As the fingerprint function, \aenet, and thus abICS, support symmetry functions \cite{behler2007,behler2011,behler2015} and Chebyshev descriptors \cite{artrith2017}. In the abICS scheme, the same model is used to map a configuration on an ideal lattice to the energy after relaxation, so that the total energy after relaxation may be expressed as 
\begin{equation} \label{eq:latNNP}
    E_\text{rel}(\conf) = \sum_i^\text{atoms} \text{NN}_{t_i}^\text{rel} (f[\conf_i^{R_\mathrm{c}}]) 
    \text{ for } \conf \in \{\conf_\text{lattice}\},
\end{equation}
where $E_\text{rel}(\conf)$ represents the energy after structural
relaxation from a starting structure $\conf$, 
$f[\conf_i^{R_\text{c}}]$ represents atomic environment descriptors for
the starting structure, and $\text{NN}_{t_i}^\text{rel}$ represents atomic
energy contributions to the relaxed total energy. The training and use of the
model is restricted to $\conf \in \{\conf_\text{lattice}\}$, where
$\{\conf_\text{lattice}\}$ represents the set of ideal lattice structures
with configurational disorder.

An important point to note is the option, or sometimes the necessity, to ignore one or more species in the on-lattice model. In the case of a two-component alloy with fixed composition, it is possible to train the on-lattice model to predict the energy based on the occupation of only one of the components. It is also advisable to ignore a component when that species has the same occupation of sites in all considered configurations; a typical example is when considering disorder only in the cation sublattice in multi-component oxides. Ignoring a component is often also necessary because machine learning usually starts by normalizing the variances of descriptors. This means that if all atoms of a certain species has the same local environment, one can end up with NaN (Not a Number) errors due to zero variance.

\subsection{Monte Carlo method} \label{sec:MC}
One of the main features of abICS is to generate atomic configurations obeying the canonical distribution $p(\conf; \beta) = \exp(-\beta E(\conf)) / Z(\beta)$ at a given temperature $T$, where $\conf$ denotes a configuration, $E(\conf)$ is the energy of the configuration, $\beta = 1/T$ is the inverse temperature, and $Z(\beta) = \sum_{\conf} p(\conf; \beta)$ is the partition function.
The numerator $W(\conf; \beta) = \exp(-\beta E(\conf))$ is called the Boltzmann factor.

The Markov chain Monte Carlo (MCMC) method generates a series of configurations $\set{\conf_t}$ efficiently. ($t$ denotes an index of MC steps.)
In the MCMC method, a trial configuration $\conf'$ is proposed with a transition probability $T(\conf\rightarrow \conf')$. The trial configuration is chosen by slightly modifying the current one $\conf = \conf_t$, for example, by exchanging the positions of two atoms. Then, the candidate configuration is accepted with the acceptance probability $p(\conf \rightarrow \conf')$, $\conf_{t+1} = \conf'$.
If rejected, $\conf_t$ survives as the next configuration, $\conf_{t+1} = \conf_{t}$.
When the acceptance probability $p(\conf \rightarrow \conf')$ satisfies the balanced condition, $W(\conf') = \sum_{\conf} W(\conf) T(\conf\rightarrow \conf') p(\conf\rightarrow \conf')$, the distribution of the generated configurations $\{\conf_t\}$ converges to the desired distribution, $p(\conf) = W(\conf)/Z$ after a sufficiently large number of configurations are generated.~\footnote{The configurations generated in the early stage of the simulation should be discarded because they may not obey the target distribution. This is called the thermalization or burn-in process.}
The expectation value of an observable $A$ at the inverse temperature $\beta$ can be obtained simply by taking the average over the generated configurations;
\begin{equation}
\braket{A}_\beta = \frac{1}{Z(\beta)}\sum_{\conf} A(\conf) W(\conf; \beta) \simeq \frac{1}{N_\text{MC} - N_\text{th}} \sum_{t = N_\text{th}+1}^{N_\text{MC}} A(\conf_t),
\end{equation}
where $N_\text{MC}$ is the number of the whole MC steps and $N_\text{th}$ is the number of the burn-in steps.

To calculate the proper acceptance probability, the Metropolis-Hasting method
\begin{equation}
p(\conf \rightarrow \conf')
= \min\left[1, \frac{W(\conf')T(\conf'\rightarrow \conf)}{W(\conf)T(\conf\rightarrow \conf')}\right]
= \min\left[1, e^{-\beta (E(\conf') - E(\conf))} \frac{T(\conf'\rightarrow \conf)}{T(\conf\rightarrow \conf')}\right]
\end{equation}
is usually used.
In the symmetric case, $T(\conf\rightarrow \conf') = T(\conf'\rightarrow \conf)$, the acceptance probability becomes $p(\conf\rightarrow \conf') = \min\left[1, \exp(-\beta (E(\conf')-E(\conf)))\right]$.
This means that the MCMC method can escape from the valley (local minima) with an energy depth $\Delta E \sim T$.
In other words, the MCMC method tends to get stuck in local minima, particularly at low temperatures.

In the past few decades, several advanced algorithms have been developed to overcome this problem, and the extended ensemble algorithm is a general one among them.
This algorithm extends the configuration space and/or modifies the distribution, for example, by changing the temperature as a configuration parameter.
At present, abICS implements two types of extended ensemble algorithms: The replica exchange Monte Carlo (RXMC) method~\cite{Swendsen1986, Hukushima1996} and the population annealing Monte Carlo (PAMC) method~\cite{Neal2001, Hukushima2003}.
Both algorithms suit massively parallel computers, although it has been suggested that the PAMC method is more suitable than RXMC method for massively parallel implementations \cite{wang_comparing_2015}.
We briefly introduce them in the following.

\subsubsection{Replica exchange Monte Carlo method}
In the RXMC method~\cite{Swendsen1986, Hukushima1996}, $N$ MCMC random walkers (replicas) at  different inverse temperatures $\beta_1, \beta_2, \dots, \beta_{N}$ run in parallel.
At preset intervals, the temperatures are exchanged according to the following procedure;
a replica $i$ with the energy $E_i$ and the inverse temperature $\beta_i$ and another replica $j$ with $E_j$ and $\beta_j$ swap the temperatures with the probability
\begin{equation}
p = \min\left[1, \frac{
e^{-\beta_i E_j} e^{-\beta_j E_i}
}{
e^{-\beta_i E_i} e^{-\beta_j E_j}
} \right]
=
\min\left\{1, 
e^{(\beta_i-\beta_j)(E_i-E_j)}
\right\}.
\end{equation}
Focusing on one replica, the temperature of the replica can increase and decrease through the simulation, and hence the replica that is trapped in a local minimum has a chance to escape when the temperature becomes increased.
Since the exchange probability is derived by the Metropolis-Hastings algorithm, after a sufficiently large number of MC steps, the exchange operation does not distort the distribution, and therefore the generated configurations with temperature $\beta_i$ obey the canonical distribution $p(\conf; \beta_i)$.

\subsubsection{Population annealing Monte Carlo Method}
% ToDo: check
The population annealing Monte Carlo (PAMC) method~\cite{Hukushima2003} extends the simulated annealing (SA) method.
In SA, the MCMC method starts at a high temperature and the temperature is gradually lowered.
This simple procedure is widely used to search for the configuration with the lowest energy, that is, the ground state.
The temperature-decreasing process, however, leads to a distortion in the distribution of random walkers, deviating from equilibrium.
This is why the SA method is not suitable for generating configurations at each temperature; for that, a burn-in process would be required for every temperature increment.

To address this issue, the annealed importance sampling (AIS) method~\cite{Neal2001} is proposed.
The AIS prepares $N$ independent replicas and runs the SA method on these replicas with $\beta=0$ in parallel.
The AIS method introduces additional weights $w_i$ called Neal-Jarzynskii (NJ) weights to each replica $i$ in order to compensate for the distortion of the distribution.
The initial value of $w_i$ is 1, and on decreasing the temperature from $\beta$ to $\beta' > \beta$, $w_i$ is updated as
\begin{equation}
    w_i \leftarrow w_i \frac{W(\conf_i; \beta')}{W(\conf_i; \beta)} = w_i e^{-(\beta' - \beta) E(\conf_i)}.
\end{equation}
By using the NJ weights, the expectation value of an observable $A$ under the inverse temperature $\beta$ is the weighted average over the replicas;
\begin{equation}
\braket{A}_\beta = \frac{1}{Z(\beta)}\sum_{\conf} A(\conf) W(\conf; \beta) \simeq \frac{1}{N} \sum_{i}^{N} A(\conf_i) w_i.
\end{equation}

Since replicas with lower energies tend to have larger NJ weights, the variance of the NJ weights increases as the simulation progresses particularly when the energy landscape has many local minima.
It reduces the number of effective samples and results in poor computational accuracy.
The PAMC method mitigates this problem by introducing resampling based on the NJ weights of the replicas.
After each resampling, all NJ weights are reset to unity.
This resampling step helps stuck replicas escape from local minima and reduces the variance of the weights. 

\subsubsection{Free energy}
The free energy $F$, which is connected with the partition function as $Z = e^{-\beta F}$, is a crucial quantity to investigate the stability of the model because nature favors the model with lower free energy.
While the partition function itself is difficult to calculate by the MC method,
the ratio of the partition functions between two temperatures can easily be calculated as
\begin{equation}
\begin{split}
  \frac{Z(\beta')}{Z(\beta)}
  =
  \frac{\sum_{\conf} \exp\left[-\beta' E(\conf)\right]}{\sum_{\conf} \exp\left[-\beta E(\conf)\right]}
  &=
  \frac{\sum_{\conf} \exp\left[-(\beta'-\beta) E(\conf)\right]
  \exp\left[-\beta E(\conf)\right]
  }{\sum_{\conf} \exp\left[-\beta E(\conf)\right]} \\
  &=
  \Braket{e^{-(\beta'-\beta)E}}_\beta.
  \end{split}
  \label{eq:Zdiff}
\end{equation}
The partition function in the high-temperature limit ($\beta=0$)  is just the volume of the configuration space $\Omega$, and hence the partition function at other temperatures can be calculated as $Z(\beta) = \Omega Z(\beta)/Z(0)$.
Unfortunately, the accuracy of the Eq.(\ref{eq:Zdiff}) becomes worse as the difference between temperatures becomes larger.
This is because the energy range contributing to the partition function depends on temperature; generally, this is a small region around the expectation value of energy $\braket{E}_\beta$.
To overcome this problem, Eq.~\eqref{eq:Zdiff} is evaluated iteratively as
\begin{equation}
  Z(\beta_N) 
  =
  \Omega\frac{Z(\beta_N)}{Z(\beta_1=0)}
  =
  \Omega
  \frac{Z(\beta_N)}{Z(\beta_{N-1})}
  \frac{Z(\beta_{N-1})}{Z(\beta_{N-2})}
  \cdots
  \frac{Z(\beta_{2})}{Z(\beta_{1}=0)},
\end{equation}
where $\beta_N > \beta_{N-1} > \cdots > \beta_1 = 0$.
The RXMC method and the PAMC method generate the configurations for several temperature points, and thus this algorithm can be easily applied in these methods.

\subsection{Representation of the structure} \label{sec:repst}
\begin{figure}[tbp]
    \centering
    \includegraphics[width=12cm]{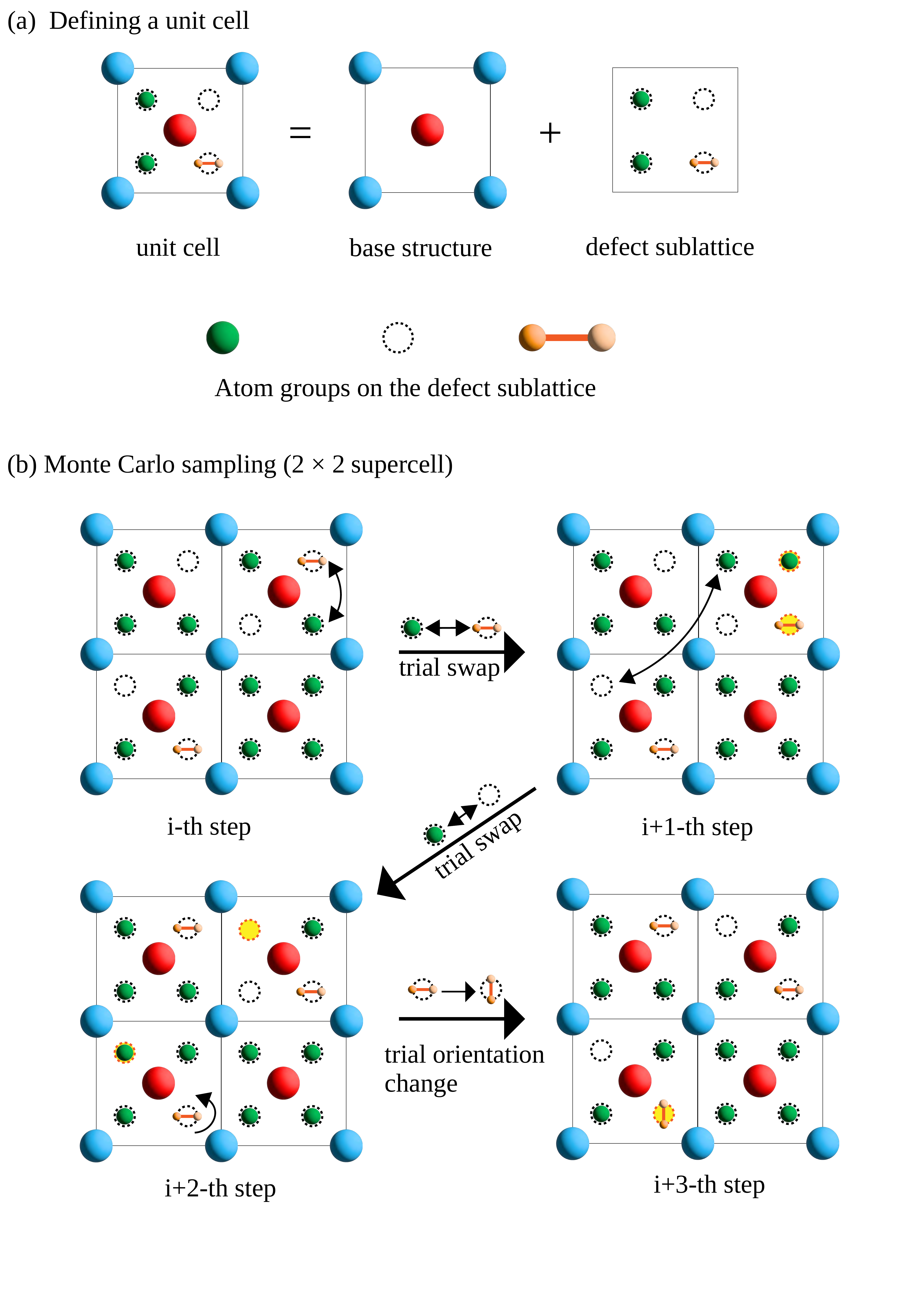}
    \caption{(a) Schematic figure of a unit cell comprised of a base structure and one defect sublattice, where three atom group types (a green atom, a vacancy, and an orange molecule) can occupy sites on the defect sublattice. (b) A schematic of Monte Carlo sampling on a $2 \times 2$ supercell of (a). The yellow shaded circles indicate updated sites.}
    \label{fig:configs}
\end{figure}

The basic unit cell that defines the ideal lattice with configurational disorder can be separated into a ``base structure'' whose atom occupations are held fixed during MC sampling and one or more ``defect sublattices'' on which MC sampling is performed (Fig.~\ref{fig:configs}(a)). Internally, a defect sublattice is represented by an array of site coordinates along with
a list of ``atom group'' types that reside on each site. In abICS, an atom group can be a single atom, a group of atoms (e.g., a molecule residing on a lattice site) or a vacancy.
Multiple sublattices with different sets of atomic species can be defined. For example,
it is natural and often more efficient to define cation and anion sublattices for ionic solids. It is also possible to
define a single sublattice with anion and cation mixing, but in many systems, anions residing on
cation sites and vice versa are unstable and contribute very little to the thermodynamics.
% With the above information, it is trivial to convert the internal representation to input structure files for
% first-principles or machine learning potential solvers.
% In abICS, we use pymatgen's \texttt{Structure} object \cite{Ong2013} as a common intermediate representation that is fed into wrappers for the solvers (\aenet, LAMMPS, Quantum ESPRESSO, VASP, and OpenMX).

The internal representation of sublattices described above enables easy implementation of configuration sampling on a lattice.  For random configuration generation in the initial step of active learning, the atom group list of each defect sublattice is randomly shuffled and converted to input for first-principles calculation code. In Monte Carlo sampling, trial atom swaps or orientation changes of atom groups (Fig.~\ref{fig:configs}(b)) are performed by swapping elements of the atom group list and converting to input for the configuration energy calculator. 

% The definition of the configuration in abICS is outlined using Fig.~\ref{fig:configs} as an example.
% Figure~\ref{fig:configs}~(a)--(c) schematically depicts unitcell, base\_structure, and defect\_structure, where red and blue circles are the atomic types defined by base\_structure, respectively. The dotted circles indicate the location of the defects defined by defect\_structure, which accommodate atom groups denoted by green circles or may be left vacant.
% Figure~\ref{fig:configs}~(d) shows the atomic species in base\_structure. Here, two atomic species of red and blue are defined. 
% Figure~\ref{fig:configs}~(e) is a schematic figure for specifying the group of atoms to be located at the defect position of defect\_structure.
% The green circle defines a group that consists of one atom of one species, and the orange circles form a group of two atoms composed of two atom species. A group may have an orientational degree of freedom. 
% The defect\_structure and the atom groups that belong to the defect sublattice can be defined multiple times.

% Figure~\ref{fig:configs}~(f) schematically describes the construction of trial configurations during the Monte Carlo sampling. There are two types of canonical \textit{moves}: One is to rotate the orientation of one randomly chosen element of atom groups. The other is to swap the location of two random elements of distinct species that reside in the same defect sublattice.

\subsection{Structure mapping}
The success of a lattice model depends on a good mapping between the structures before and
after relaxation. However, in ionic systems especially with vacancies, atoms sometimes relax out of their initial sites
into adjacent sites, breaking the one-to-one mapping mentioned above. That is, the relaxed structure sometimes does not
correspond to the internal representation that was used to generate the initial structure for relaxation. In this case, it is more appropriate to remap the relaxed structure to the lattice \cite{barroso-luque_cluster_2022}. To achieve this, abICS uses a simple distance-based mapping for each sublattice. The remapped structures on the rigid lattice and the corresponding relaxed energies are then used for training the configuration energy model.

\subsection{Constraints for avoiding inaccessible configurations} \label{sec:inaccessible}
Relaxation out of initial sites mentioned above implies that there are lattice representations that are inaccessible after relaxation \cite{barroso-luque_cluster_2022}. These inaccessible configurations may still be sampled by Monte Carlo algorithms, which can be an issue because no training data can be accumulated for such configurations. If the energy model happens to predict a low enough energy for such configurations, they will be sampled and the resulting thermodynamics will be distorted to varying degrees. On the other hand, if it is possible to predict inaccessible configurations from, e.g., chemical intuition, one can impose constraints on the structures that are accepted by Monte Carlo sampling and reject the inaccessible configurations. abICS provides a functionality to impose such constraints.

\section{Installation}
\label{sec:installation}
abICS is implemented in Python3 (version 3.7 or later), and relies on the following external Python libraries: NumPy \cite{harris_array_2020}, SciPy \cite{2020SciPy-NMeth}, toml \cite{toml}, mpi4py \cite{dalcin_mpi4py_2021}, pymatgen (version 2019.12.3 or later) \cite{Ong2013}, and qe-tools \cite{qe-tools}. For installing mpi4py, which handles parallel calculations, the user needs to have a working message-passing interface (MPI) library installed in the system.

Since abICS has been registered in Python Package Index (PyPI) repository, users can install abICS easily by the following command:

\begin{verbatim}
  $ python3 -m pip install abics
\end{verbatim}
If the installation should be done locally, e.g. when the user has no permission to install files outside of their home directory, \texttt{--user} option may be added to the command.

The source code of abICS is available from GitHub repository. Users can download the zipped source archive from the release site, or clone the repository by
\begin{verbatim}
  $ git clone https://github.com/issp-center-dev/abICS.git
\end{verbatim}
After unpacking the source archive and moving into the top of the source tree, users can install abICS by typing
\begin{verbatim}
  $ python3 -m pip install .
\end{verbatim}
The installation directory may be explicitly specified by \texttt{--prefix=DIRECTORY} option in which \texttt{DIRECTORY} is the path to the directory where abICS will be installed.

In addition, \aenet has to be installed for neural network training and evaluation. We provide interfaces to \aenet via either a file I/O-based method (write out \aenet input, run \aenet as a subprocess, and parse the output file) or through a LAMMPS \cite{thompson2022} Python interface which doesn't rely on file I/O. The latter is optional but is generally faster in the MC sampling stage managed by \texttt{abics\_sampling}. To enable the Python interface, the user must install aenet-LAMMPS \cite{chen2021} as well as the LAMMPS Python module. It is also necessary to have a working installation of Quantum Espresso, VASP, and/or OpenMX for calculating the reference training data.

% \subsection{Postprocessing (Motoyama)}
% In abICS, \verb|abicsRXsepT| is provided as the postprocessing tool of replica exchange Monte Carlo run.
% This tool is for reordering the resulting structures and energies at each sampling step of a RXMC run by temperature. It is used after an abICS RXMC run is finished as:
% \begin{verbatim}
% $ mpiexec -np NPROCS abicsRXsepT /
%   input.toml NSKIP
% \end{verbatim}
% \verb|NPROCS| should be equal to or larger than the number of replicas, and \verb|input.toml| should be replaced by the abICS input file that was used for this run. \verb|NSKIP| is an optional parameter and is used for specifying the number of thermalization steps to skip when calculating the energy averages at each temperature (default value is 0). The results are stored in the \verb|Tseparate| directory, and energy averages vs. temperature are stored in \verb|Tseparate/energies_T.dat|.

%Add example of Tseparate/energies_T.dat

%----------------------------------------------------------------
\section{Format of input and output files}
\label{sec:fileformat}

%Add schematic figure of abICS
\subsection{Input file format}
\label{sec:fileformat:input}
The overall active learning procedure is controlled by an input file in TOML format, an easy-to-read configuration format used in many software projects. The parameters for the first principles calculations (e.g., $k$-point sampling, plane wave energy cutoff, relaxation algorithm, etc.) and the neural network training/evaluation (neural network structure and various hyperparameters) are controlled by separate files following the formats of the specified calculation code. We will refer to the  files as ``reference input files''. Also, any program that calculates physical properties such as the energy from a configuration is referred to as a ``solver''; this includes first-principles calculation codes and neural network evaluation codes.

The input parameter file of abICS consists of five blocks, called \textit{sections} hereafter, that classify the parameters in categories with labels enclosed by brackets.
In the following, we briefly describe the specifications of each section. Further details can be 
found in the online manual \cite{abics-manual}.

% - - - - - - - - - - - - - - - - - - - - - - - - - - - - - - - -
\subsubsection{\texttt{\upshape [sampling]} section}
\label{sec:fileformat:input:sampling}
The sampling section specifies the parameters for Monte Carlo sampling:
for example, when using the RXMC method, the number of replicas, the temperature range,
the number of Monte Carlo steps, the replica exchange trial frequency, etc. are given. 
It contains the \texttt{[sampling.solver]} subsection that specifies the type of energy calculator (referred to as ``solver'' hereafter) used in the sampling. When performing active learning, \aenet is the only choice, but abICS also partially supports direct combination of first-principles solvers (Quantum Espresso, VASP, and OpenMX) with MC sampling. In addition, the path to the solver and the directory containing reference input files are specified. 
%It also accepts \texttt{user} type to use user-defined functions for evaluating the energy of a given structure.
An example of \texttt{[sampling]} section for RXMC sampling with 8 replicas is shown as follows (please see the online manual for PAMC parameters):
\begingroup
\setlength{\listingindent}{0pt}
\vskip2ex\hrule
\begin{listing}
[sampling]
sampler = "RXMC"
nreplicas = 8
nprocs_per_replica = 1 
kTstart = 600.0
kTend = 2000.0
nsteps = 6400
RXtrial_frequency = 4
sample_frequency = 16
print_frequency = 1
reload = false
\end{listing}
\hrule\vskip2ex
\endgroup
\noindent
An example of \texttt{[sampling.solver]} section is shown as follows:
\begingroup
\setlength{\listingindent}{0pt}
\vskip2ex\hrule
\begin{listing}
[sampling.solver]
type = "aenet"
path= "predict.x"
base_input_dir = "./baseinput"
perturb = 0.0
run_scheme = "subprocess"
ignore_species = ["O"]
\end{listing}
\hrule\vskip2ex
\endgroup
\noindent
In this example, \texttt{type} is set to \aenet in order to use the neural network model.
\texttt{path} specifies the path to \texttt{predict.x} of \aenet library that estimates the value of energy corresponding to the structure.
The directory that contains the reference input file is given by \texttt{base\_input\_dir}.
\texttt{perturb} is a parameter for randomly shifting atomic positions prior to structural optimization. This is useful for avoiding high-symmetry saddle points when relaxing the system but should be set to zero as in this example when using the on-lattice model.
\texttt{run\_scheme} specifies the way to invoke the solver. In this example, abICS starts \texttt{predict.x} using subprocesses.
\texttt{ignore\_species} specifies the atomic species to \textit{ignore} in on-lattice models as discussed in Sec.~\ref{sec:olmodel}.

% - - - - - - - - - - - - - - - - - - - - - - - - - - - - - - - -
\subsubsection{\texttt{\upshape [mlref]} section}
\label{sec:fileformat:input:mlref}
The \texttt{mlref} section specifies options applied when the atomic configurations are retrieved from the results of Monte Carlo calculations. These parameters are used, for example, to evaluate the accuracy of the neural network models, or to extend the training data.
An example of \texttt{[mlref]} is shown as follows:
\begingroup
\setlength{\listingindent}{0pt}
\vskip2ex\hrule
\begin{listing}
[mlref]
nreplicas = 8
ndata = 5
\end{listing}
\hrule\vskip2ex
\endgroup

It contains a subsection named \texttt{[mlref.solver]} to specify the parameters for the solver used to generate training data. The format is the same as that of \texttt{[sampling.solver]}. An example is shown below:
\begingroup
\setlength{\listingindent}{0pt}
\vskip2ex\hrule
\begin{listing}
[mlref.solver]
type = "qe"
base_input_dir = "./baseinput_ref"
perturb = 0.05
ignore_species = []
\end{listing}
\hrule\vskip2ex
\endgroup
\noindent
In this example, \texttt{type} is set to \texttt{qe} in order to use Quantum ESPRESSO.
\texttt{path} specifies the path to \texttt{pw.x}, the self-consistent field (SCF) energy solver of Quantum ESPRESSO.
The directory that contains the input parameter files specific to each solver is given by \texttt{./baseinput\_ref}.
\texttt{perturb} is set to $0.05$ to apply shifting atomic positions prior to structural optimization in units of angstrom.
\texttt{ignore\_species} is set to an empty list to take all species into account during the first-principles calculations. 

% - - - - - - - - - - - - - - - - - - - - - - - - - - - - - - - -
\subsubsection{\texttt{\upshape [train]} section}
\label{sec:fileformat:input:train}
The parameters for \texttt{abics\_train} are specified in the \texttt{[train]} section. An example is shown as follows:
\begingroup
\setlength{\listingindent}{0pt}
\vskip2ex\hrule
\begin{listing}
[train]
type = "aenet"
base_input_dir = "./aenet_train_input"
exe_command = ["generate.x", "srun train.x"]
ignore_species = ["O"]
\end{listing}
\hrule\vskip2ex
\endgroup

% - - - - - - - - - - - - - - - - - - - - - - - - - - - - - - - -
\subsubsection{\texttt{\upshape [observer]} section}
\label{sec:fileformat:input:observer}

The \texttt{[observer]} section contains the settings for the physical observables to be measured. It contains subsections referring to the specific types of observables.
Examples of \texttt{[observer]} subsections are given as follows:
\begingroup
\setlength{\listingindent}{0pt}
\vskip2ex\hrule
\begin{listing}
[observer]

[observer.similarity]
reference_structure = "./MgAl2O4.vasp"
ignored_species = ["O"]

[[observer.solver]]
name = "obs1"
type = "aenet"
path= "predict.x"
base_input_dir = "./baseinput_obs"
perturb = 0.0
run_scheme = "subprocess"
ignore_species = ["O"]

[[observer.solver]]
name = "coordnum"
type = "user"
function = "mymodule.coordnum"
\end{listing}
\hrule\vskip2ex
\endgroup
In the above example, two types of observables are shown: \texttt{[observer.similarity]} and \texttt{[observer.solver]}.
The former specifies options for evaluating the \textit{similarity} of atomic configuration, defined by the ratio of atoms of each element to be found at the same place as those of the reference state.
The latter specifies the physical quantities to be evaluated by using the solver.
In this example, the first \texttt{[observer.solver]} means that an observable, ``obs1'' of the configuration is calculated by using the \aenet model. A practical example of using this functionality is when a separate on-lattice model is trained to predict physical properties such as the lattice volume \cite{hoshino2023}.
The second \texttt{[observer.solver]} means that another observable, ``coordnum'', is calculated by using a Python function \texttt{mymodule.coordnum} defined in the following file named \texttt{mymodule.py}\footnote{Because abICS searches the current working directory as well as the standard places for Python packages, this file should be put in the directory where \texttt{abics\_sampling} is executed.},

\begingroup
\setlength{\listingindent}{0pt}
\vskip2ex\hrule
\begin{listing}
from pymatgen.core import Structure

def coordnum(st: Structure) -> float:
    # write code for calculating, e.g., 
    # coordination number from pymatgen structure 
    # and substitute result into n
    ...
    return n
\end{listing}
\hrule\vskip2ex
\endgroup

% - - - - - - - - - - - - - - - - - - - - - - - - - - - - - - - -
\subsubsection{\texttt{\upshape [config]} section}
\label{sec:fileformat:input:config}

\texttt{[config]} section specifies the base structure and one or more defect sublattices explained in Sec.~\ref{sec:repst}. 
The contents of this section are briefly described as follows:
\begin{itemize}
\item
  \texttt{[config]} specifies the atomic position. The simulation cell (in units of angstrom) is given by \texttt{unitcell} and \texttt{supercell}.
  Constraints on the structure to avoid inaccessible configurations (Sec.~\ref{sec:inaccessible}) can be imposed by the optional \texttt{constraint} parameter that specifies a user-defined function to determine whether a given structure is allowed or not. 
\item
  \texttt{[[config.base\_structure]]} specifies the atoms that are left fixed during the Monte Carlo calculations.
  \texttt{type} specifies the atom species, and \texttt{coords} specifies the fractional coordinates of the atoms.
  \texttt{coords} is a list of three-component arrays. It can be given by a string that represent a numeric matrix with three columns separated by white spaces and as many rows as the number of atoms.
\item
  \texttt{[[config.defect\_structure]]} specifies the position of the atoms that move during the Monte Carlo calculation.
\item
  \texttt{[[config.defect\_structure.groups]]} specifies the name and number of atoms or atom groups to be moved in the MC calculation. 
\end{itemize}

An example of \texttt{[config]} section is as follows:
\begingroup
\setlength{\listingindent}{0pt}
\vskip2ex\hrule
\begin{listing}
[config]
unitcell = [
     [8.1135997772, 0.0000000000, 0.0000000000],
     [0.0000000000, 8.1135997772, 0.0000000000],
     [0.0000000000, 0.0000000000, 8.1135997772]]
supercell = [1,1,1]
constraint = "constraint_module.constraint_func"

[[config.base_structure]]
type = "O"
coords = [
     [0.237399980, 0.237399980, 0.237399980],
     [0.762599945, 0.762599945, 0.762599945],
        ... Skipped ...,
     [0.262599975, 0.262599975, 0.762599945],
     ]
     
[[config.defect_structure]]
coords = [
     [0.000000000, 0.000000000, 0.000000000],
     [0.749999940, 0.249999985, 0.499999970],
        ... Skipped ...,
     [0.124999993, 0.624999940, 0.124999993],
     ]
     
[[config.defect_structure.groups]]
name = "Al"
species = ["Al"]
coords = [[[0,0,0]]]
num = 16

[[config.defect_structure.groups]]
name = "Mg"
species = ["Mg"]
coords = [[[0,0,0]]]
num = 8
\end{listing}
\hrule\vskip2ex
\endgroup
\noindent
In this example, abICS will sample the possible configurations of 16 \texttt{Al} and 8 \texttt{Mg} atoms on the lattice specified by \texttt{[[config.defect\_structure]]}.
The constraint on the configurations is imposed by the function \texttt{constraint\_func} defined in the file \texttt{constraint\_module.py} which takes a pymatgen Structure object and returns True if the structure is acceptable and false if it should be rejected.  A trivial example of the constraint function reads as follows:
\begingroup
\setlength{\listingindent}{0pt}
\vskip2ex\hrule
\begin{listing}
from pymatgen.core import Structure

def constraint_func(st: Structure) -> bool:
    return True
\end{listing}
\hrule\vskip2ex
\endgroup
\noindent This constraint functionality was used to constrain the position of interstitial protons to be next to occupied oxygen sites in Ref.~\cite{hoshino2023}.

Since writing all the coordinates by hand is quite tedious, we also provide a preprocessing tool \texttt{st2abics} to convert structure formats supported by pymatgen into a template TOML file with the 
\texttt{config} section filled in.

%----------------------------------------------------------------
\subsection{Output files}
\label{sec:fileformat:output}

The calculation process of abICS consists of alternating executions of training data generation steps, training steps, and Monte Carlo sampling.
The progress of steps is recorded in \texttt{ALloop.progress}.
The output files generated during these steps are summarized in this section.

In the training data generation steps, \texttt{abics\_mlref} first populates the \texttt{AL}\textit{n} directory with subdirectories containing input files and initial structures for first-principles solvers, where \textit{n} stands for the active learning iteration count staring from zero.
\texttt{abics\_mlref} also generates a list of these directories in \texttt{rundirs.txt}, and the user is tasked with running first-principles solvers in those directories. After the solvers are run, the results are postprocessed by \texttt{abics\_mlref} for the next training step. Also, the correspondence between the previous on-lattice model prediction and the first-principles solver results are written in files named \texttt{energy\_corr.dat}  for validation of the on-lattice energy model.

In the training steps using \texttt{abics\_train}, the input and output files are stored in a directory \texttt{train0}. The generated on-lattice model is placed in the input directory specified in \texttt{[sampling.solver]}.

In the Monte Carlo sampling steps using \texttt{abics\_sampling}, the input and output files are stored in a directory \texttt{MC}\textit{n} where \textit{n} corresponds to the index of the preceding AL step.
When the calculation of \texttt{MC}\textit{n} step finishes, the results are outputted into files of several types.
They are summarized as follows.
For each MC replica, the files listed below are generated in the individual replica directory:
\begin{itemize}
\item \texttt{structure.NNN.vasp}: The atomic coordinates for each step are stored in the POSCAR file format of VASP. Here, \texttt{NNN} corresponds to the MC sampling step.
\item \texttt{minE.vasp}: The structure having the lowest energy among the samples in this replica is stored in POSCAR format.
\item \texttt{obs.dat}: The temperature and the total energy for each step is shown in units of eV.
\item \texttt{obs\_save.npy}: The energy in units of eV at each step is written in the Numpy array format.
\item \texttt{kT\_hist.npy}: The temperature in units of eV at each step is written in the Numpy array format.
\item \texttt{Trank\_hist.npy}: The history of the temperature index along the steps is written in the Numpy array format (only for RXMC).
\item \texttt{logweight\_hist.npy}: The logarithm of the Neal-Jarzynski weight at each step is written in the Numpy array format (only for PAMC).
\item \texttt{acceptance\_ratio.dat}: The acceptance ratio of the Monte Carlo sampling at each temperature is summarized in the text format.
\end{itemize}

The measured physical observables are written to the files with their labels as filenames. They contain the values of temperature, the average and the statistical error.
In \texttt{logZ.dat} file, the logarithms of the partition function $\log Z_i/Z_0$ with $i$ the index of temperature are written.

%----------------------------------------------------------------

% \section{Calculation of Physical Quantities}
% After the Monte Carlo sampling procedure, the expectation value and standard error of each physical quantity 
% \texttt{abics_sampling}

\section{Applications}
Here, we review some recent applications of abICS to demonstrate the power of the approach.
\subsection{Degree of Inversion in Spinel oxides}

\begin{figure}[htb]
    \centering
    \includegraphics[width=\textwidth]{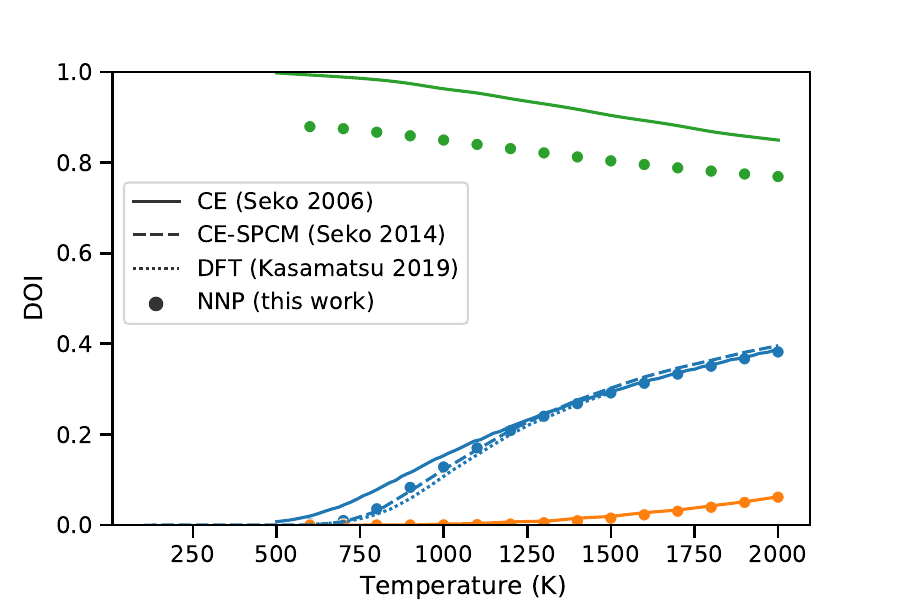}
    \caption{The degree of inversion calculated using abICS on a 192-cation cell compared to the cluster expansion literature \cite{Seko2006, Seko2014} and direct RXMC sampling in combination with
    first-principles calculation in a 48-cation cell \cite{Kasamatsu_2019}.}
    \label{fig:spinelDOI}
\end{figure}

\begin{figure}[htb]
    \centering
    \includegraphics[width=\textwidth]{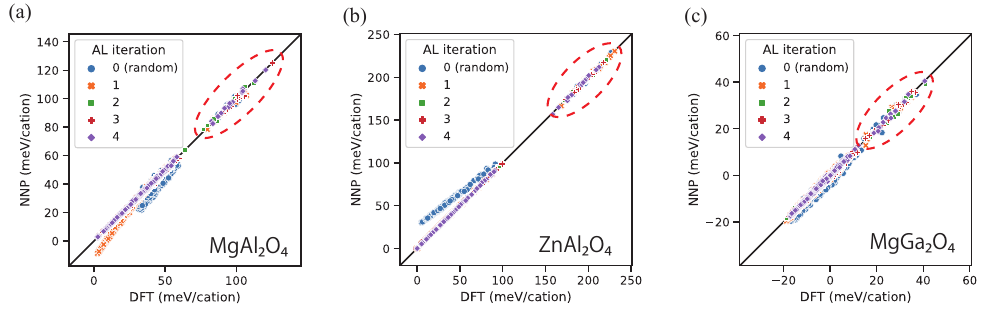}
    \caption{Improvement in the NN energy predictions vs.~DFT energies with the number of active learning iterations for (a) \ce{MgAl2O4}, (b) \ce{ZnAl2O4}, and (c) \ce{MgGa2O4}. }
    \label{fig:ALvsDFT}
\end{figure}

Monte Carlo sampling of the lattice configuration problem is naturally suited for predicting order-disorder transitions. Thus, as a first demonstration of the abICS active-learning approach, we calculated the degree of cation disorder in \ce{A^{2+}B^{3+}2O4} spinel oxides \ce{MgAl2O4}, \ce{ZnAl2O4}, and \ce{MgGa2O4} \cite{Kasamatsu2022}. 
In an ordered spinel, the divalent A cation is tetrahedrally coordinated by oxygen, while the trivalent B cation is octahedrally coordinated. With increasing temperature, there can be some degree of inversion (DOI) between the two sites, which is defined as the ratio of tetrahedral sites occupied by B cations. There are also some A/B combinations that result in almost complete inversion, and \ce{MgGa2O4} is one of them. 

Fig.~\ref{fig:spinelDOI} shows the calculated DOI compared with some cluster expansion literature \cite{Seko2006,seko2011,Seko2014} and direct RXMC sampling on first-principles energies \cite{Kasamatsu_2019}. The results are in excellent agreement except for some quantitative difference in the DOI of \ce{MgGa2O4}, which we speculate is due to the small unit cell used for training the cluster expansion model. The improvement in the accuracy of the neural network configuration energy model with the number of active learning cycles is shown in Fig.~\ref{fig:ALvsDFT}. The prediction errors decrease to less than 1 meV/cation using less than 1000 DFT samples of the 192-cation cell, which can be calculated within a few hours on a modern supercomputer system. 
We refer the reader to Ref.~\cite{Kasamatsu2022} for further details including analysis of the influence of lattice relaxation, training set cell size, and a detailed comparison with random training set generation.

\subsection{Hydration of \ce{Sc}-doped \ce{BaZrO3}}

\begin{figure}[htb]
    \centering
    \includegraphics[width=\textwidth]{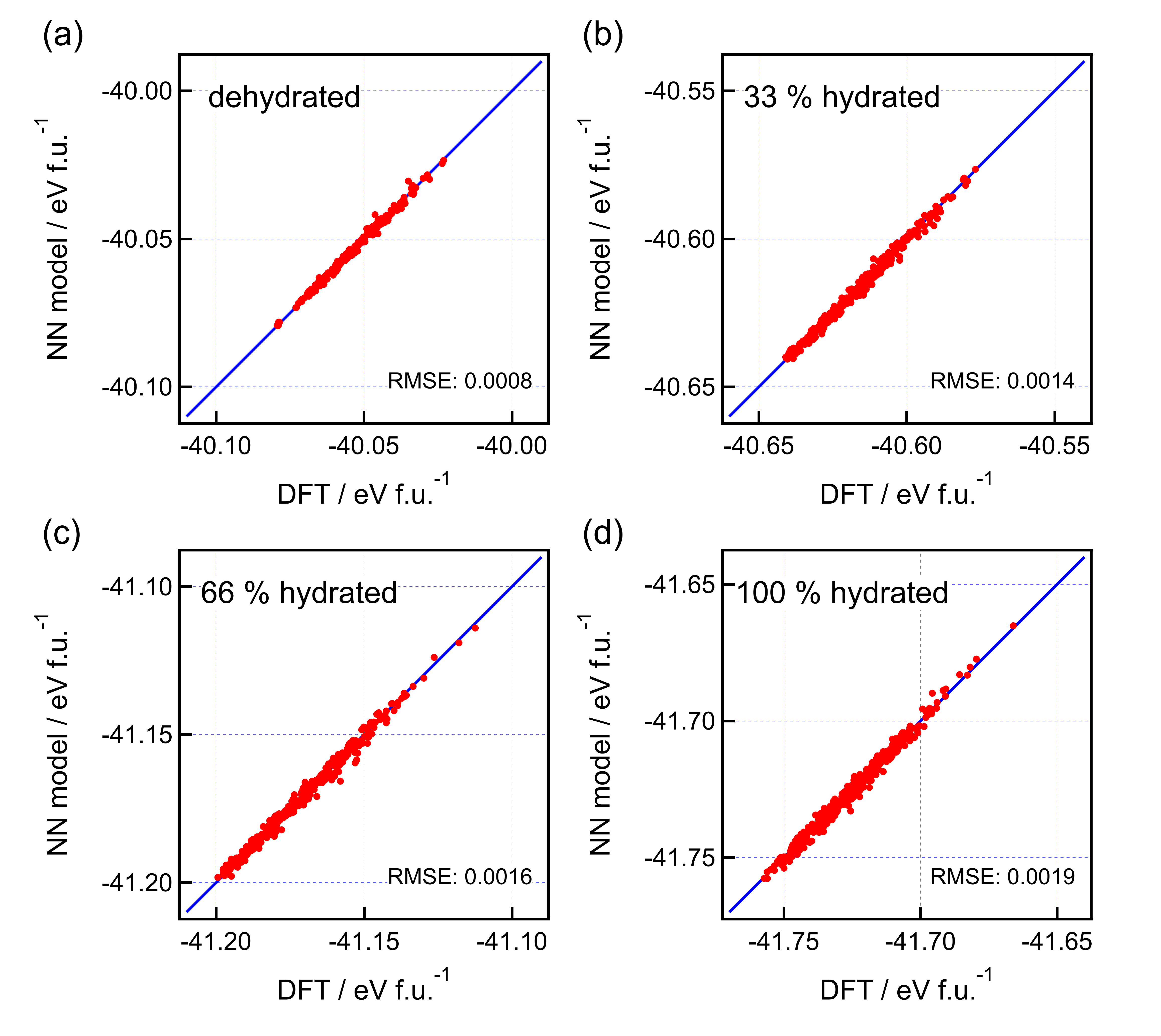}
    \caption{The NN energy model prediction vs.~DFT energies in 22.2 at\% \ce{Sc}-doped \ce{BaZrO3} with varying levels of hydration. Reproduced from Ref.~\cite{hoshino2023} with permission from the American Chemical Society.}
    \label{fig:BZScNN}
\end{figure}

\begin{figure}[htb]
    \centering
    \includegraphics[width=\textwidth]{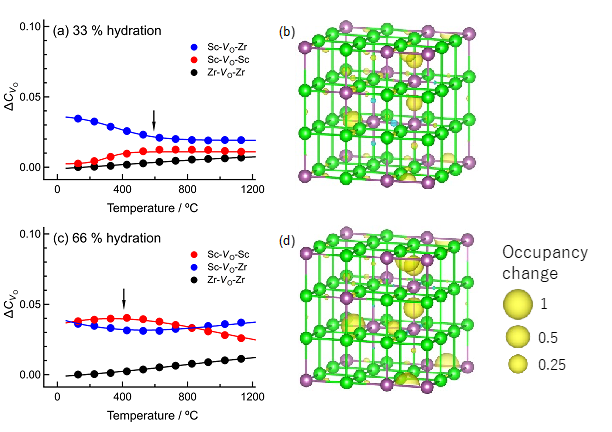}
    \caption{Concentration of oxygen vacancies filled upon hydration at (a) 33\% and (c) 66\% in 22.2 at\% \ce{Sc}-doped \ce{BaZrO3}, and the occupancy change of each oxygen site at (b) 33\% and (c) 66\% hydration denoted by the size of the yellow spheres. Only the \ce{Zr} (green)/\ce{Sc} (purple) sublattice is shown in (b) and (d). Reproduced from Ref.~\cite{hoshino2023} with permission from the American Chemical Society.}
    \label{fig:BZScdC}
\end{figure}

The spinel DOI problem was chosen as our first demonstration because it was beneficial to have a benchmark result using a conventional method, i.e., cluster expansion. 
More recently, we have been applying abICS to much more difficult systems that would be virtually impossible to treat using cluster expansion. 
One such system is acceptor-doped \ce{BaZrO3}, which is known as a promising proton conductor for fuel cell applications\cite{hyodo_fast_nodate,hyodo_accelerated_2021}.

Doping of a trivalent cation such as Y and Sc on the tetravalent Zr site induces the formation of oxide ion vacancies to satisfy charge neutrality. 
These oxygen vacancies act as active centers for the hydration reaction under humid atmosphere, which results in the filling of oxygen vacancies and introduction of mobile protons into the system. 
It is well known that there is a strong sample dependence of the amount of water uptake in this class of materials, and researchers have been trying to understand this behavior in the hopes of optimizing the proton carrier concentration. 

A hypothesis is that there are active and inactive oxygen vacancies depending on the vacancy environment, and the number of each environment would depend on the dopant type and processing conditions; 
when considering the nearest neighbor site, there are three types of oxygen vacancies in the Sc-doped case: Sc-V$_\mathrm{O}$-Sc, Sc-V$_\mathrm{O}$-Zr, and Zr-V$_\mathrm{O}$-Zr. 
From the experimental side, probing of the changes in the local environment upon hydration requires careful {\it in situ} measurements, and a clear picture is yet to be attained. 
From the theoretical side, it may seem straightforward to perform Monte Carlo calculations to simulate which vacancy types are filled as hydration proceeds, but this is very challenging because of the complexity of the material; 
Sc and Zr reside on the Zr site, O or V$_\mathrm{O}$ reside on the O site, and protons or proton vacancies can be introduced in interstitial positions, 
making this a multi-lattice six-component configuration problem that is virtually impossible to handle using conventional methods. In Ref.~\cite{hoshino2023}, we used abICS to tackle this issue. 

Fig.~\ref{fig:BZScNN} shows the extremely high prediction accuracy of our on-lattice neural network model for 22 at\% \ce{Sc}-doped \ce{BaZrO3} with varying levels of hydration. 
The model was used to compute the hydration behavior as a function of hydration levels and temperature (Fig.~\ref{fig:BZScdC}), and careful comparison with {\it in situ} X-ray absorption and thermogravimetry measurements led to the following clarification: 
Water intake with decreasing temperature under constant vapor pressure starts with the filling of Sc-V$_\mathrm{O}$-Zr sites; 
once these sites are almost completely filled, Sc-V$_\mathrm{O}$-Sc sites start to be occupied. 
Zr-V$_\mathrm{O}$-Zr, which was predicted to be the most active from hydration enthalpy calculations, were shown to exist in negligible amounts under realistic equilibrium conditions and contributes very little to the hydration behavior. 

%\subsection{AuLi GC?}

\subsection{Interfacial space charge at solid-state electrochemical interfaces}
Another topic that we have been applying abICS to is the issue of the interfacial space charge at solid-solid interfaces.
The concept of space charge is well known in semiconductor physics where a layer of electrons or holes is formed to align the Fermi level across interfaces. The concept can be extended to include ionic defects, which are often the dominant charge carriers when considering interfaces in, e.g., solid oxide fuel cells or solid state Li-ion batteries. Recent years have seen increasing interest in understanding and experimentally probing such effects in energy conversion devices \cite{maier2005,ohta2006,chen_interface_2021,banerjee2020}, but success has been limited to very few state-of-the-art electron microscopy techniques \cite{nomura_direct_2019}. From the theory side, phenomenological modeling based on the Poisson-Boltzmann model has provided some insight \cite{Kasamatsu2011,Kasamatsu2012, Shimizu2020,swift_modeling_2021}. However, explicit atomistic modeling has been limited to examination of position-dependent defect formation energies in the dilute limit \cite{Kasamatsu2010,Haruyama2014}, and calculations at realistic carrier concentrations are lacking. This is because efficient thermodynamic sampling requires a lightweight effective model, and conventional models such as cluster expansion are very difficult to converge at surfaces and interfaces \cite{Yuge2007}. Our preliminary calculations show that the on-lattice neural network model combined with high-throughput supercomputing may provide a solution.

\begin{figure}[htb]
    \centering
    \includegraphics[width=\textwidth]{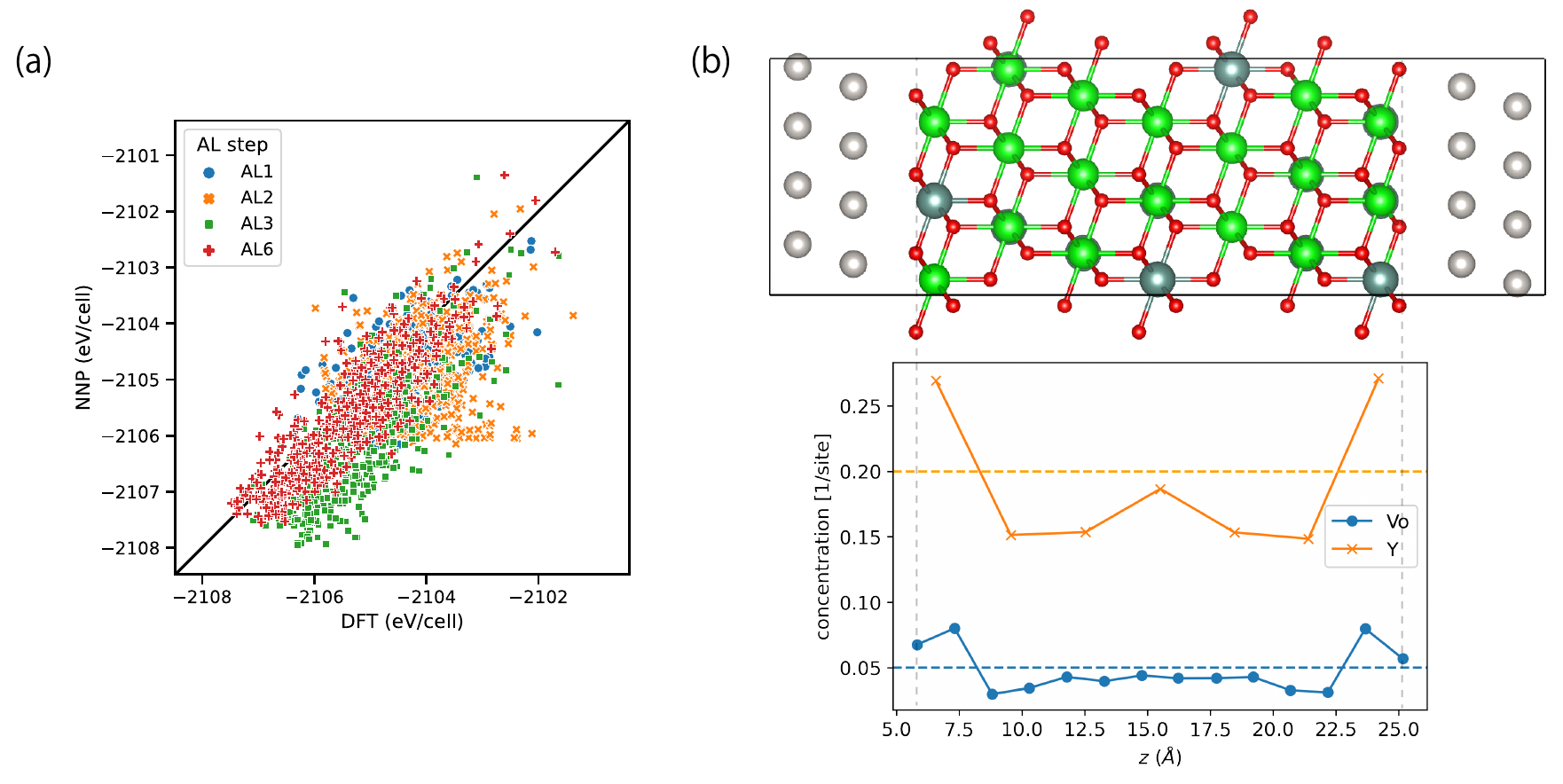}
    \caption{(a) Improvement of the on-lattice neural network model with active learning iterations and (b) Y and O vacancy concentration profiles in the calculated Pt/YSZ slab model.}
    \label{fig:PtYSZ}
\end{figure}

Fig.~\ref{fig:PtYSZ} shows the accuracy of the on-lattice model and calculated ion concentration profiles for a Pt (111)/\ce{Y2O3}-doped \ce{ZrO2} (yttria stabilized zirconia, or YSZ) slab model. YSZ is known as a well-studied oxide ion conductor for solid oxide fuel cell applications. The parent lattice is a cubic fluorite \ce{ZrO2} with 68 Zr sites and 126 O sites, and configurations of 12 Y ions on Zr sites and 6 O vacancies residing on the O sites were sampled. The Pt slab contains 64 atoms. Although the attained accuracy of the on-lattice neural network model is not as good as the bulk systems introduced above, the root mean square error is $< 2.5$ meV/atom if we count all atoms (Pt, Zr, O, Y) and 
$<3.2$ meV/atom if we only count the oxide atoms (Zr, O, Y), which is considered small enough to obtain sufficiently accurate thermodynamics. The calculated concentration profile at 1800 K, which corresponds to the sintering temperature, indicates segregation of Y dopants and O vacancies at the interface. The next challenge would be to consider the variation of O vacancy concentrations under varying oxygen partial pressures, which provide the chemical driving force for fuel cell operation. The free energy and grand canonical calculation methods implemented recently in abICS will enable such calculations.

\section{Summary and Outlook}
In this paper, we introduced the open source Python framework abICS for first-principles based statistical thermodynamics of lattice systems. A robust on-lattice machine learning energy model is trained against first-principles data through active learning iterations combined with parallel statistical sampling methods (RXMC or PAMC). This enables calculations on many-component many-sublattice systems and their interfaces, which are beyond the reach of conventional methods.

We stress that abICS is still under heavy development. As already mentioned, grand canonical sampling is currently under testing. We are also considering the integration of various machine learning models which have been under heavy
development in the past few years. Suggestions for new functionalities are welcome, and the modular object-oriented code structure should enable
other developers to implement new functionalities and algorithms. 
With these developments, we are certain that abICS will contribute greatly to expanding the scope of first-principles materials modeling for understanding of the thermodynamics of complex materials. 

\section*{Acknowledgments}
The development of abICS has been supported by various funding bodies and computer resource providers over the years.
\begin{itemize}
    \item ``Priority Issue on Post-K computer'' (Development of new fundamental technologies for high-efficiency energy creation, conversion/storage and use)
    \item ``Program for Promoting Research on the Supercomputer Fugaku'' (Fugaku Battery \& Fuel Cell Project JPMXP1020200301 and Fugaku Materials Physics \& Chemistry Project JPMXP1020230325)
    \item Leading Initiative for Excellent Young Researchers (LEADER) by Ministry of Education, Culture, Science, and Technology (MEXT) of Japan
    \item KAKENHI (No. 15K20953, JP18H05519, 19K15287, 20H05284, 21H01041, 22H04606) by MEXT 
    \item Core Research for Evolutional Science and Technology (CREST) by Japan Science and Technology Agency (No. JPMJCR15Q3)
    \item New Energy and Industrial Technology Development Organization
    \item Fusion Oriented Research for disruptive Science and Technology (FOREST) by Japan Science and Technology Agency (No. JPMJFR2037)
    \item Joint-use Supercomputer System at The Institute for Solid State Physics, The University of Tokyo
    \item Supercomputer Fugaku provided by the RIKEN Center for Computational Science (Project ID: hp230205).
\end{itemize}
We would also like to express our thanks for support from ``Project for advancement of software usability in materials science'' of The Institute for Solid State Physics, The University of Tokyo.

% References
\bibliographystyle{tfnlm}
\bibliography{abics}

\end{document}